\documentclass[]{article}
\usepackage[a4paper, total={6in, 9in}]{geometry}
\usepackage{mathtools}
\usepackage{amsmath}
\usepackage{caption}
\usepackage{float}
\usepackage{subcaption}
\usepackage{graphicx}
\usepackage{bibref}
\usepackage{tabularray}
\usepackage{multirow}
\title{Observational Evidence to Logistic Dark Energy Driving the Accelerating Universe}
\author{Sarath Nelleri\footnote{sarathnelleri@gmail.com}, Gopi Krishna\footnote{pndktz91199@gmail.com} and Navaneeth Poonthottathil\footnote{navaneeth@iitk.ac.in}  \\ \small $^2$Department of Physics, Indian Institute of Technology, Kanpur 208016, India}
\date{}
\begin{document}

\maketitle

\begin{abstract}
We present logistic dark energy model (LDEM), where the dark energy density follows a logistic function for the scale factor. The equation of state parameter of dark energy ($w_D$) transitioned from $-1$ in the distant past to its current value of $-0.76$, closely resembling the $\Lambda$CDM model in the early epoch and showing significant deviation in the late phase.  The evolution of the deceleration parameter in the LDEM signifies its success in explaining the late-time cosmic acceleration. Model selection based on the Bayesian Information Criterion (BIC), incorporating observations from Type Ia Supernovae (SNe Ia), Observational Hubble data (OHD), and Baryon Acoustic Oscillation (BAO) strongly favors the LDEM over the conventional $\Lambda$CDM model, where BIC is estimated to be $\sim -20$. Incorporating the shift parameter derived from the Cosmic Microwave Background (CMB) data shows competing evidence of the LDEM over the standard $\Lambda$CDM. Remarkably, the Hubble constant ($H_0$) value computed using any of the datasets tends to align closely with the predictions from the Cosmic Microwave Background (CMB), suggesting a need to reconsider the local measurement.
\end{abstract}

\section{Introduction}
Accelerated expansion of the universe was solidly established through comprehensive observational probes, including Type Ia supernovae (SNe Ia) \cite{perlmutter1999measurements, riess1998observational, perlmutter1999constraining}, Cosmic Microwave Background Radiation (CMB) \cite{mather1990preliminary, mather1994measurement, tegmark2003high, hinshaw2013nine, ade2016planck, aghanim2020planck}, Baryon Acoustic Oscillation (BAO) \cite{beutler20116df, aubourg2015cosmological}, Large Scale structure (LSS) in the distribution of galaxies \cite{efstathiou1992cobe, tegmark2004cosmological, alam2017clustering, beutler20116df} and direct measurement of redshift and distances \cite{ma2011power, hernandez2020generalized, sudharani2023hubble}. The $\Lambda$CDM scenario has become the widely accepted cosmological model due to its simplicity and its concordance with various cosmological observations. Observations of the Cosmic Microwave Background Radiation (CMB) and local measurements of cosmological parameters across various redshifts reveal a statistically significant discrepancy \cite{di2021realm, schoneberg2019bao+, aloni2022step, alestas2021late, mortsell2018does}. For instance, the estimation of $H_0$ by the Planck collaboration 2018 predict a value of $67.4\pm 0.5$ km $\text{s}^{-1}$ $\text{Mpc}^{-1}$\cite{aghanim2020planck} whereas, the observation on Cepheids by the SHOES collaboration yielded a value of $74.03\pm 1.42$ km $\text{s}^{-1}$ $\text{Mpc}^{-1}$\cite{riess20162, riess2021cosmic}. This discrepancy highlights the tension in the precise measurement of the Hubble constant, which stands at approximately 
$4.2\sigma$. Here, the Planck findings are relying on the $\Lambda$CDM model, while the local measurements are made without assuming any specific cosmological model. The highly precise measurements from the Cosmic Microwave Background (CMB) data strongly support the standard cosmological model. Researchers typically attribute the observed discrepancies to the $\Lambda$CDM model, rather than considering the possibility of systematic uncertainties in the local measurements. However, both possibilities remain viable. These observed disparities may shed light on the most compelling evidence of physics beyond the standard cosmology or the need to reconsider the local measurement. 
Various efforts have been made to resolve these challenges. For instance, ref. \cite{poulin2019early} presents early dark energy that behaves like a cosmological constant in the early universe, alleviating the Hubble tension. Other notable examples comprise phantom dynamical dark energy models\cite{dahmani2023smoothing}, negative dark energy models\cite{sen2023cosmological, malekjani2023negative}, interacting dark matter dark energy scenario \cite{gariazzo2022late} the dissipative axion particle model\cite{berghaus2020thermal}, baryon inhomogeneities resulting from primordial magnetic fields\cite{jedamzik2020relieving}, modified gravity models\cite{adi2021can, adil2021late}, Omnipotent dark energy \cite{adil2024omnipotent}, sign switching cosmological constant \cite{akarsu2021relaxing, akarsu2023relaxing} and many more. See ref. \cite{di2021realm} for a comprehensive review of various attempts to resolve the Hubble tension. Nevertheless, certain models leave the discrepancy well above $3\sigma$, while others alleviate the Hubble tension due to larger error bars in the estimated value of $H_0$. Some models that mitigate the Hubble tension exhibit dynamic instability, and support for other models, in contrast to the $\Lambda$CDM, is limited when considering the observational data. Recent studies suggest that the Hubble tension problem involves an evolution of the $H_0$ with redshift.\cite{malekjani2023negative, dainotti2022quasar, dainotti2022evolution}.

To address the Hubble tension problem, we introduce a parameterized dark energy model named the Logistic Dark Energy Model (LDEM), where the evolution of dark energy mimics a logistic function of the scale factor. A comprehensive analysis based on observational data such as SNe Ia, OHD, BAO, and CMB shows strong evidence supporting LDEM over the conventional $\Lambda$CDM model. Interestingly, the value of $H_0$ predicted by LDEM closely matches that of $H_0$ obtained using high-redshift data across all data combinations considered.

The paper is organized as follows. In Sec. 2, we introduce the logistic dark energy model (LDEM). In Sec. 3, we perform Bayesian inference to constrain the model parameters and present a detailed model comparison based on the Bayesian Information Criterion (BIC). In sec. 4, we study the evolution of cosmographic parameters, estimate its present value, and compare it with the standard model prediction. Finally, we conclude in sec. 5.

\section{Logistic dark energy model (LDEM)}
\begin{figure}
	\centering
	\includegraphics[width=0.6\textwidth]{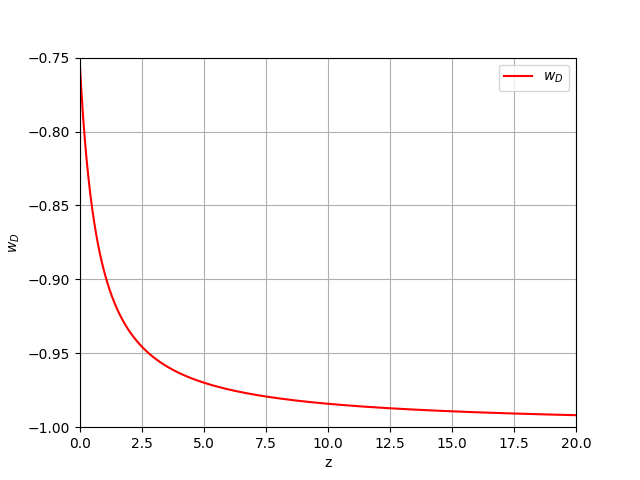}
	\caption{Evolution of equation of state parameter of dark energy ($w_D$) against redshift ($z$).}
	\label{fig:1}
\end{figure}

We introduce the logistic dark energy model, where dark energy density follows a simple logistic function in scale factor, aiming to explain the late-phase acceleration of the universe. We consider the logistic dark energy density (normalized over the present critical density $\rho^0_c$), defined as
\begin{align}
	\label{eqn:1}
	\Omega_D(a) = \frac{\alpha}{e^{a} + 1},
\end{align} 
where $a$ is the scale factor and $\alpha$ is a dimensionless constant. To satisfy $\Omega_D = \Omega^0_D$ at $a=1$, with $\Omega^0_D$ denoting the present value of the dark energy density parameter, we determine $\alpha$ as $(e+1)\Omega^0_D$. The functional form presented in Eq. (\ref{eqn:1}) holds particular interest for two reasons. First is its frequent occurrence in physics, notably resembling the Fermi-Dirac distribution and the Woods-Saxon nuclear density distribution. The second reason lies in the statistical simplicity of the LDEM model, sharing an equal number of parameters with the well-established $\Lambda$CDM model. Certainly, our primary motivation for this study is rooted in the second reason – the simplicity of the LDEM model and the prospect of obtaining stronger evidence in light of observational data compared to the concordance $\Lambda$CDM model.
Any novel cosmological model must account for the late-phase acceleration of the universe. The requisite condition for the accelerated expansion is that the equation of state of dark energy, $w_D$, must be less than -1/3. Assuming no interaction between dark energy and non-relativistic matter, both components adhere to the conservation equation\cite{baumann2022cosmology},
\begin{align}
	\label{eqn:2}
	\dot{\Omega_i} + 3H(1+w_i)\Omega_i = 0,
\end{align}
where $i=m, D$ represents matter and dark energy. Considering pressureless matter with $w_m = 0$, the evolution of matter density is expressed as,
\begin{align}
	\label{eqn:3}
	\dot{\Omega_m} = \Omega^0_{m} a^{-3}.
\end{align}
Using the Eq. (\ref{eqn:1}) and (\ref{eqn:2}) for dark energy, we obtain the evolution of $w_D$ as 
\begin{align}
	\label{eqn:4}
	w_D = -1 + \frac{1}{3}\left[\frac{ae^a}{e^a + 1}\right].
\end{align}
The evolution of $\omega_D$ with the scale factor is illustrated in Fig. (\ref{fig:1}).
The plot demonstrates $\omega_D$ evolving from $-1$ at $z\rightarrow \infty$ to $w_D = -1+[e/(3(e+1))]$ at present ($z=0$). The current value of $w_D$ is $-0.756$, indicating the potential for an accelerating expansion of the universe. The evolution of the universe is characterized by the Friedmann equation\cite{baumann2022cosmology},
\begin{align}
    \label{eq:fried}
    H^2 = \frac{8\pi G}{3} (\rho_m + \rho_{D})
\end{align}
where we assume the curvature parameter, $k \approx 0$, and avoid the radiation density as it does not contribute significantly to the late-phase expansion rate.
The progress of the Hubble parameter with the scale factor can be obtained by substituting the evolution of dark energy and matter densities as presented in Eq. (\ref{eqn:1}), (\ref{eqn:3}) and (\ref{eq:fried}) as
\begin{align}
	\label{eq:H}
	H(a) = H_0\left(\Omega^0_m a^{-3} + \frac{(e+1)\Omega^0_D}{e^a + 1}\right)^{1/2}
\end{align}
Hubble parameter is the parameter of fundamental interest when we study the background evolution of the universe. Our subsequent interest is in constraining the model parameters and studying the evolution of cosmographic parameters.
\section{Observational constraints on model parameters}
In order to constrain the model parameters, we adopt Bayesian inference procedure. We employ recently released Type Ia supernovae data, specifically the Pantheon+ dataset (SNe Ia), along with Observational Hubble Data (OHD), Baryon Acoustic Oscillation (BAO), and shift parameter derived from the Cosmic Microwave Background (CMB) to constrain the model parameters. The Bayesian inference is based on the Bayes theorem, which in this case can be expressed as\cite{trotta2008bayes, john2002comparison, padilla2021cosmological, hobson2010bayesian, trotta2007applications},
\begin{align}
	\label{eq:B}
	P(\theta|D,M) = \frac{P(D|\theta,M)P(\theta|M)}{P(D|M)},
\end{align}
where $P(\theta|D,H)$ is the posterior distribution of the model parameter ($\theta$), $P(D|\theta,M)$ is the likelihood, $P(\theta|H)$ is the prior and $P(D|M)$ is just a normalization factor that represents the Bayesian evidence of the model. The evidence can be taken as unity without loss of generality when we explore the parameter space of a model\cite{trotta2008bayes}. However, Bayesian evidence plays a central role in the model comparison. The prior probability reflects any information available before data collection\cite{john2002comparison}. Although the choice of prior is subjective, the iterative application of Bayes theorem results in a common posterior. The likelihood represents the probability of obtaining the data given a specific model. In this context, we assume a Gaussian likelihood, defined as\cite{simon2015cfhtlens, trotta2008bayes},
\begin{align}
	\label{eq:5}
	P(D|\theta,M) \equiv \exp(-\chi^2(\theta)/2),
\end{align}
where the $\chi^2$ distribution is expressed as \cite{verde2010statistical}
\begin{align}
	\label{eq:6}
	\chi^2(\theta) = \sum_{k}^{}\left[\frac{Q_k - Q_k(\theta)}{\sigma_k}\right]^2
\end{align}
where $Q_k$ is the physical quantity obtained from the observation, $Q_k(\theta)$ is the corresponding value predicted by the model and $\sigma_k$ is the uncertainty in the measured physical quantities. The marginal likelihood of a parameter of interest, say $\theta_1$, is obtained by marginalizing over all other parameters; we obtain
\begin{align}
	\label{eq:marginal}
	p(\theta_1|D,M) = \int p(\theta|D,M)d\theta_2...d\theta_n,
\end{align}

We use the Bayesian Information Criterion ($BIC$), sometimes called the Schwarz Information Criterion, which follows from a Gaussian approximation to the Bayesian evidence in the limit of large sample size for the model selection \cite{szydlowski2015aic, trotta2008bayes}. The $BIC$ is expressed as
\begin{align}
	\label{AIC}
	BIC \equiv -2\ln \mathcal{L}_{max} + k\ln N,
\end{align}
where $L_{max}$ is the maximum likelihood value, and the term $k\ln N$ sets the penalty term for the extra number of free parameters ($k$) of the model, and $N$ is the number of data points.
\begin{figure}
	\centering
	\includegraphics[width=0.43\textwidth]{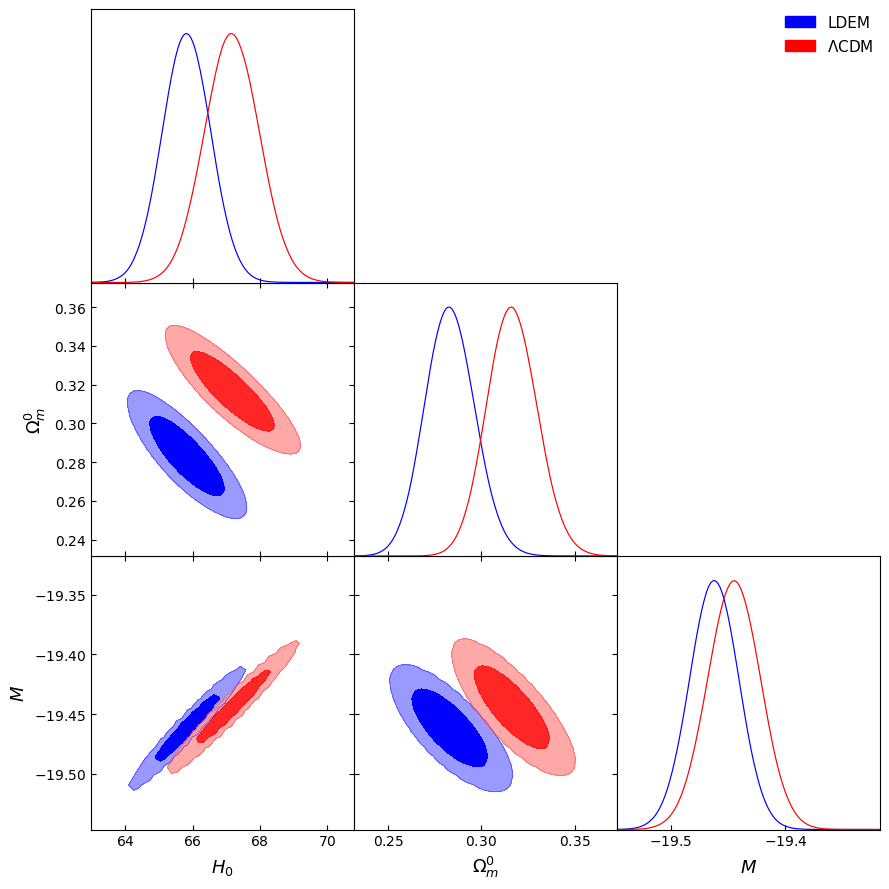}
	\includegraphics[width=0.43\textwidth]{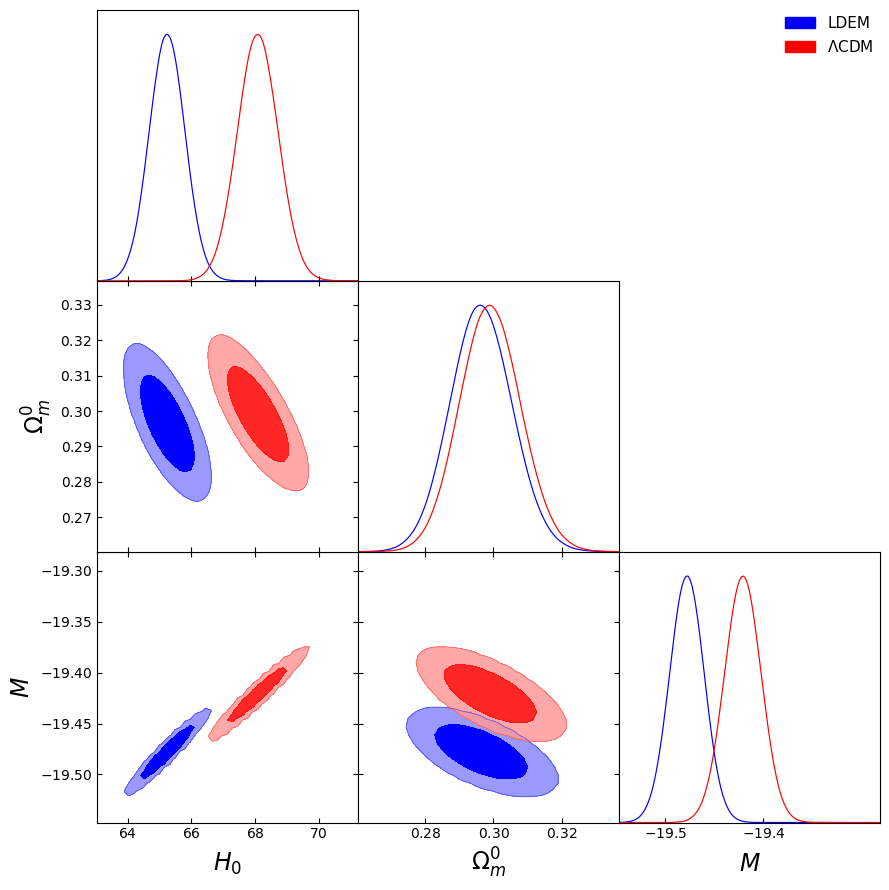}
	\caption{Comparison between the 1D and 2D
		posterior distributions of model parameters of the LDEM and $\Lambda$CDM models using the data combinations OHD+SNe Ia (Left) and OHD+SNe Ia+BAO (Right)}
	\label{fig:corner}
\end{figure}

\begin{table}
    \centering
    \caption{Comparison between $\chi^2$, BIC and the model parameters predicted by the LDEM and the $\Lambda$CDM model.}
    \resizebox{\textwidth}{!}{%
    \begin{tabular}{|c|c|c|c|c|c|c|}
        \hline
        Data & Model & $\chi^2_{\text{min}}$ & $H_0$ & $\Omega^0_m$ & $M$ & $\Delta BIC$ \\
        \hline
        OHD & LDEM & $30.19$ & $65.49^{+0.87}_{-0.89}$ & $0.2902^{+0.0185}_{-0.0175}$ & --- & $-2.23$ \\
        \cline{2-6}
         & $\Lambda$CDM & $32.42$ & $69.93^{+1.04}_{-1.06}$ & $0.2658^{+0.0171}_{-0.0160}$ & --- & \\
        \hline
        OHD + SNe Ia & LDEM & $1778.67$ & $65.82\pm 0.72$ & $0.2830^{+0.0136}_{-0.0132}$ & $-19.46\pm 0.02$ & $-20.42$ \\
        \cline{2-6}
         & $\Lambda$CDM & $1799.09$ & $67.16^{+0.82}_{-0.81}$ & $0.3163^{+0.0137}_{-0.0133}$ & $-19.44\pm 0.02$ & \\
        \hline
        OHD + SNe Ia + BAO & LDEM & $1780.53$ & $66.04\pm0.51$ & $0.2800\pm 0.01$ & $-19.46\pm0.02$ & $-20.50$ \\
        \cline{2-6}
         & $\Lambda$CDM & $1803.71$ & $68.08^{+0.64}_{-0.63}$ & $0.2990^{+0.0090}_{-0.0088}$ & $-19.42\pm0.02$ & \\
        \hline
        OHD + SNe Ia + CMB & LDEM & $1778.67$ & $65.83\pm 0.72$ & $0.2800\pm 0.01$ & $-19.46\pm 0.02$ & $-4.81$ \\
        \cline{2-6}
         & $\Lambda$CDM & $1800.04$ & $67.75\pm0.56$ & $0.3052^{+0.0071}_{-0.0069}$ & $-19.42\pm0.02$ & \\
        \hline
        OHD + SNe Ia + BAO + CMB & LDEM & $1778.67$ & $65.83\pm 0.72$ & $0.2800\pm 0.01$ & $-19.46\pm 0.02$ & $-2.70$ \\
        \cline{2-6}
         & $\Lambda$CDM & $1803.74$ & $68.02^{+0.53}_{-0.52}$ & $0.3000^{+0.0060}_{-0.0059}$ & $-19.42\pm0.02$ & \\
        \hline
    \end{tabular}%
    }
    \label{tab:chi2}
\end{table}

In the LDEM model, the number of free parameters is the same as in the $\Lambda$CDM model: the Hubble constant ($H_0$) and the present matter density parameter ($\Omega^0_{m}$). To estimate the parameters, we utilize the data sets OHD, OHD+SNe Ia, OHD+SNe Ia+BAO, OHD+SNe Ia+CMB, and OHD+SNe Ia+BAO+CMB. The OHD data set contains 57 redshift vs Hubble parameter data points, out of which 31 data points are obtained from the differential age (DA) method and 26 data points are derived from the BAO and other methods\cite{cao2021cosmological, farooq2017hubble, magana2018cardassian}. The $\chi^2$ for the OHD data can be obtained using the Eq. (\ref{eq:6}). The latest SNe Ia data set is the pantheon+ data holding 1701 light curves. In the panthon+ data, the observed physical quantity is the apparent magnitude (m) of the SNe Ia\cite{brout2022pantheon+, scolnic2022pantheon+}. The corresponding theoretical values can be computed using the expression\cite{sarath2023running}, 
\begin{align}
	\label{m}
	m_{model} = 5\log_{10}\left[\frac{d_L}{Mpc}\right] + 25 +M,
\end{align}
where $d_L$ is the luminosity distance. The cosmological model (here the LDEM) enters through the relation\cite{dainotti2022evolution, mathew2022running},
\begin{align}
	\label{dl}
	d_L(z) = c(1+z)\int_{0}^{z}\frac{dz'}{H(z)}
\end{align}
The $\chi^2$ for the pantheon+ is defined as
\begin{align}
	\label{chip}
	\chi^2_{\text{SNe Ia}} = \vec{Q}^T. (C_{stat+sys})^{-1}.\vec{Q},
\end{align}
where $\vec{Q}$ is a 1701 dimensional vector of $m_i - m_{model}(z_i)$, where $m_i$ denote the apparent magnitude of the $i^{th}$ SNe Ia and $C_{stat+sys}$ is the covariance matrix of the pantheon+ sample\cite{scolnic2022pantheon+}. We use the acoustic parameter (A) of BAO to incorporate the BAO measurements, defined as\cite{sharov2018predictions}
\begin{align}
	\label{A}
	A(z) = \frac{H_0\sqrt{\Omega^0_m}}{cz}D_V(z),
\end{align}
where $D_V(z)$ denotes the volume averaged angular diameter distance, expressed as\cite{lian2021probing}
\begin{align}
	\label{dv}
	D_V(z) = \left[\frac{czd_L^2(z)}{(1+z)^2H(z)}\right]^{1/3},
\end{align}
where $d_L(z)$ is the luminosity distance presented in Eq. (\ref{dl}).
We use 7 BAO data points in the redshift range $0.106 \leq z \leq 0.73$ in our computation\cite{alam2017clustering, sharov2018predictions}. The $\chi^2$ for the BAO is defined as
\begin{align}
	\label{bao}
	\chi^2_{\text{BAO}} = \vec{A}^T. (C_{A})^{-1}.\vec{A},
\end{align}
where $(C_{A})^{-1}$ is the covariance matrix for the BAO measurement. To incorporate CMB data, we use the shift parameter, defined as
\begin{align}
	\label{shift}
	R = \sqrt{\Omega^0_m}\int_{0}^{z_{dec}}\frac{dz'}{h(z')}
\end{align}
where $h = H/H_0$ is the reduced Hubble parameter. The observed shift parameter is $R_{obs} = 1.7502 \pm 0.0046$ at a redshift corresponding to the photon decoupling epoch, $z_{dec} = 1089.92$\cite{chen2019distance}. The $\chi^2$ for the CMB data is computed using Eq. (\ref{eq:6}). The best-estimated parameters are those that maximize the likelihood. We utilized a well-tested and open-source Python implementation of the affine-invariant ensemble sampler for MCMC, known as emcee, which was proposed by Goodman and Weare. For more details about emcee and its implementation, please see the ref.\cite{foreman2013emcee}. The 1D and 2D posterior distributions of the model parameters are presented in Fig. (\ref{fig:corner}). The mean and standard deviation of the Gaussian distribution are considered the best-fit parameter and error, respectively. The parameter values, minimum chi-square ($\chi^2_{min}$), and difference in BIC between LDEM and $\Lambda$CDM are presented in Tab. (\ref{tab:chi2}). The figures in the table are noteworthy for two reasons. Firstly, the LDEM model fits all data combinations better than the $\Lambda$CDM model, with particularly strong evidence of LDEM over the $\Lambda$CDM for the OHD+SNe Ia and OHD+SNe Ia+BAO combinations, where the $\Delta BIC$ is approximately $-20$. Note that the BIC for LDEM is comparable to $\Lambda$CDM when incorporating CMB data, which is expected because the equation of state for logistic dark energy tends towards -1 at higher redshifts. Similarly, the comparable BIC for the OHD data alone can be attributed to some data points being derived under the assumption of the $\Lambda$CDM model. Secondly, the $H_0$ value obtained from any of the data combinations predicts a value closely aligned with the CMB prediction. The LDEM model, which provides a better fit to the data than the standard model across a wide spectrum of redshifts, predicts a value close to the CMB prediction, suggesting a need to reconsider local measurements.

\section{Evolution of cosmographic parameters}
\begin{figure}
	\centering
	\includegraphics[width=0.55\textwidth]{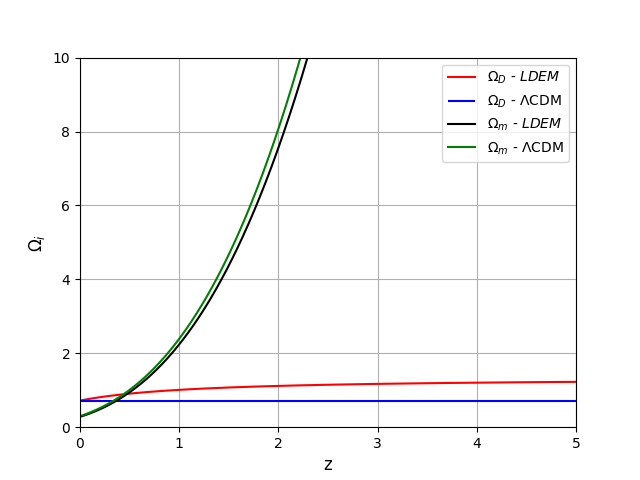}
	\caption{Evolution of matter and dark energy density normalized over the present critical density ($\Omega_i$) with redshift ($z$).}
	\label{fig:density}
\end{figure}

In this section, we study the evolution of the key cosmological parameters. The evolution of Hubble parameter is given by Eq. (\ref{eq:H}).
\begin{figure}
	\centering
	\includegraphics[width=0.55\textwidth]{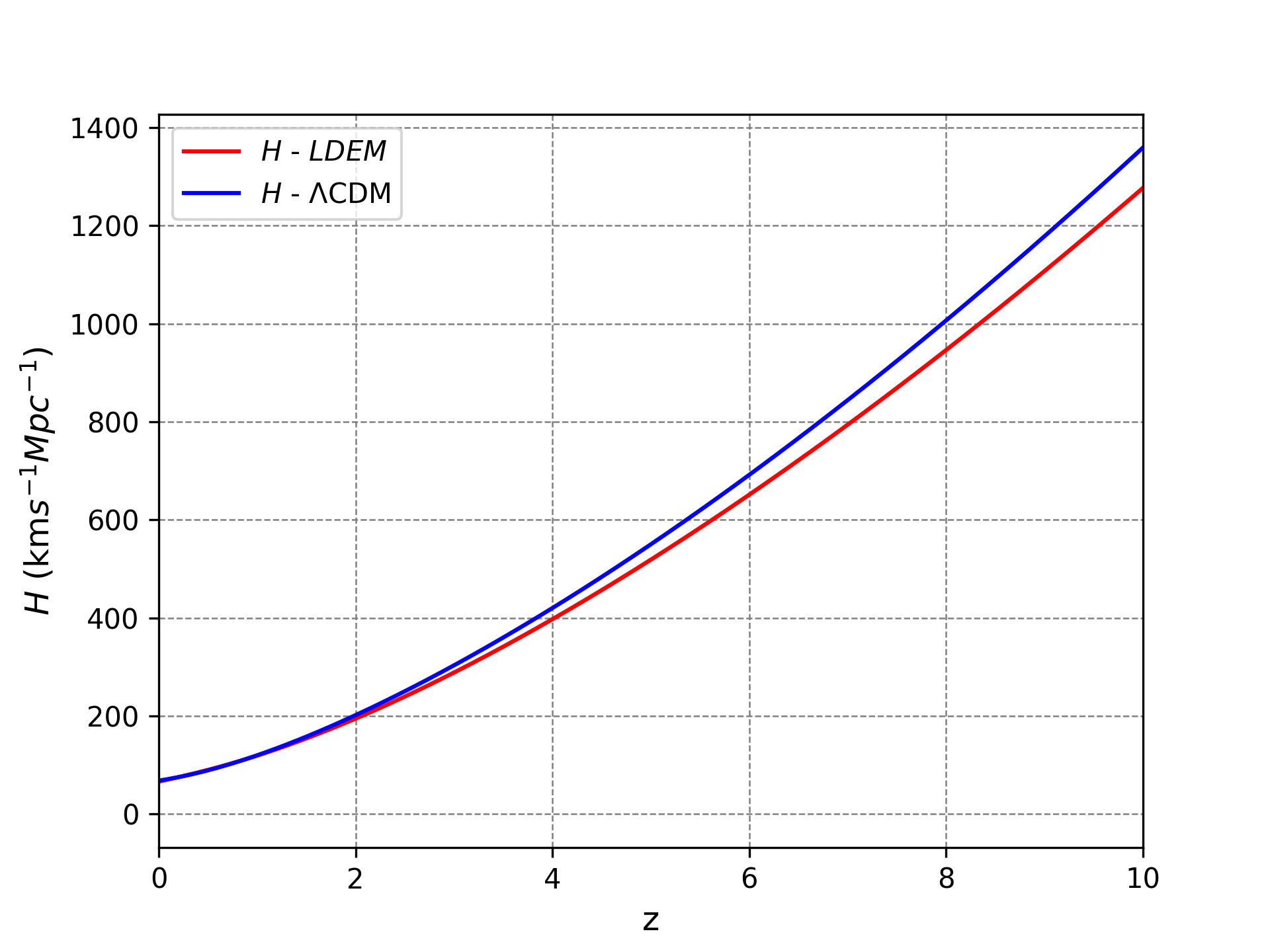}
	\caption{Evolution of Hubble parameter ($H$) with redshift ($z$)}
	\label{fig:Hubble}
\end{figure} 
\begin{figure}
	\centering
	\includegraphics[width=0.55\textwidth]{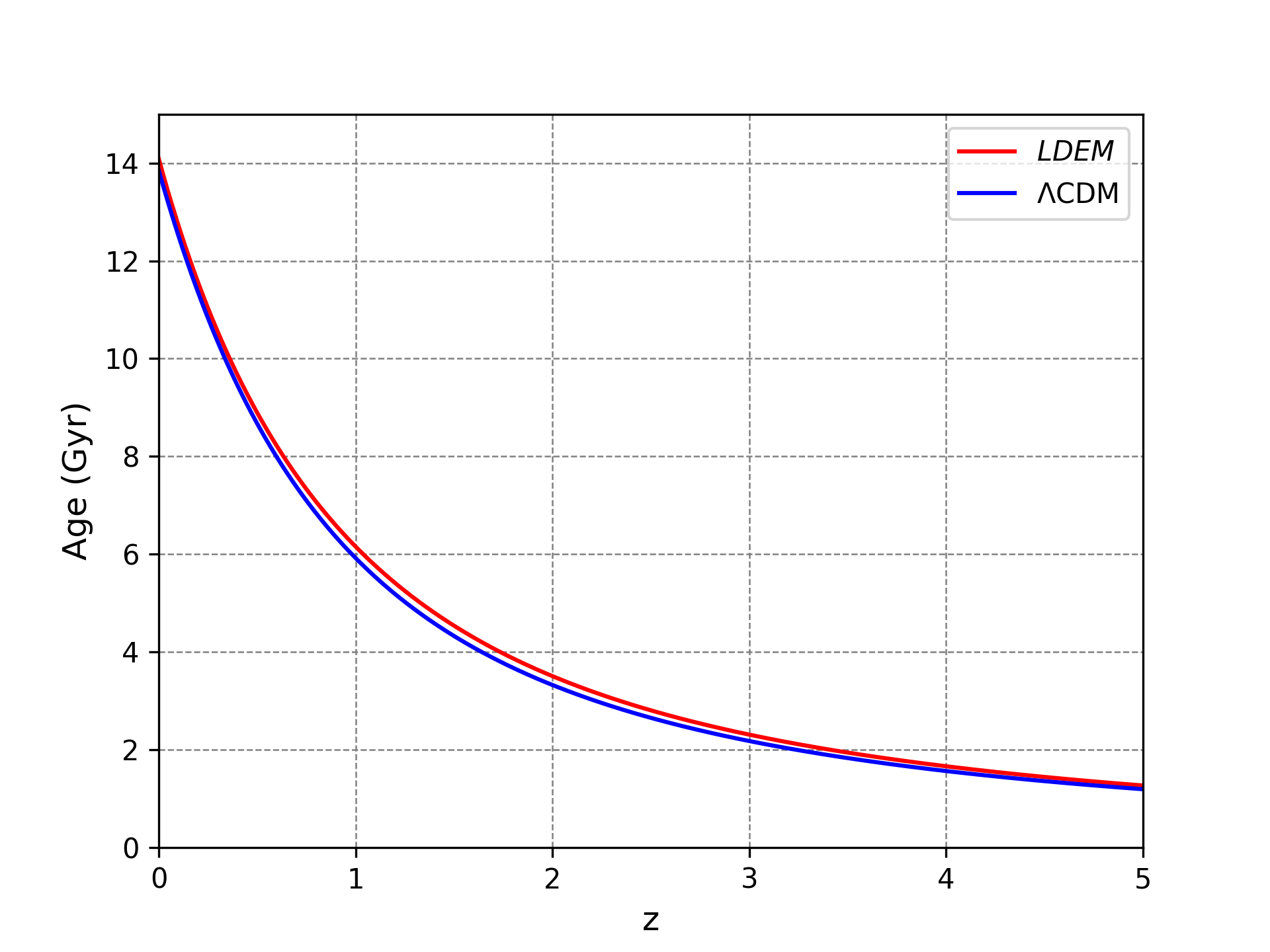}
	\caption{Progress of the age of the universe with redshift ($z$).}
	\label{fig:age}
\end{figure}
The progress of Hubble parameter with the redshift using the best-fit model parameters are shown in Fig. (\ref{fig:Hubble}). The Hubble parameter increases monotonically with redshift.
At present, it matches the value of the Hubble parameter in the $\Lambda$CDM model, but it shows significant deviations during the early matter-dominated epoch.
Present age of the universe is computed using the expression\cite{baumann2022cosmology, george2019interacting},
\begin{align}
	\label{eq:age}
	t_0 = \int_0^{1} \frac{{da}}{{a H(a)}}
\end{align}
The evolution of the age of the universe as a function of redshift is depicted in Fig. (\ref{fig:age}). 
The LDEM model predict a slightly higher age of 14.09 as compared to the age 13.85 predicted by the $\Lambda$CDM model.
Comparison between evolution of matter density and dark energy density normalized over the present critical density in LDEM and $\Lambda$CDM model are depicted in Fig. (\ref{fig:density}). The evolution of matter density in LDEM is similar to that of the $\Lambda$CDM model while the dark energy density is increasing with the redshift in LDEM model in contast to a cosmological constant in the $\Lambda$CDM model. The dark energy dominated over the matter at a redshift of 0.475 in the LDEM model while it happens at a lower redshift of 0.330 for the $\Lambda$CDM model. 

The accelerating or decelerating expansion of the universe is characterized by the deceleration parameter ($q$), defined as\cite{george2016holographic},
\begin{align}
	\label{eq:dec}
	q = -1 - \frac{\dot{H}}{H^2}.
\end{align}
\begin{figure}
	\centering
	\includegraphics[width=0.55\textwidth]{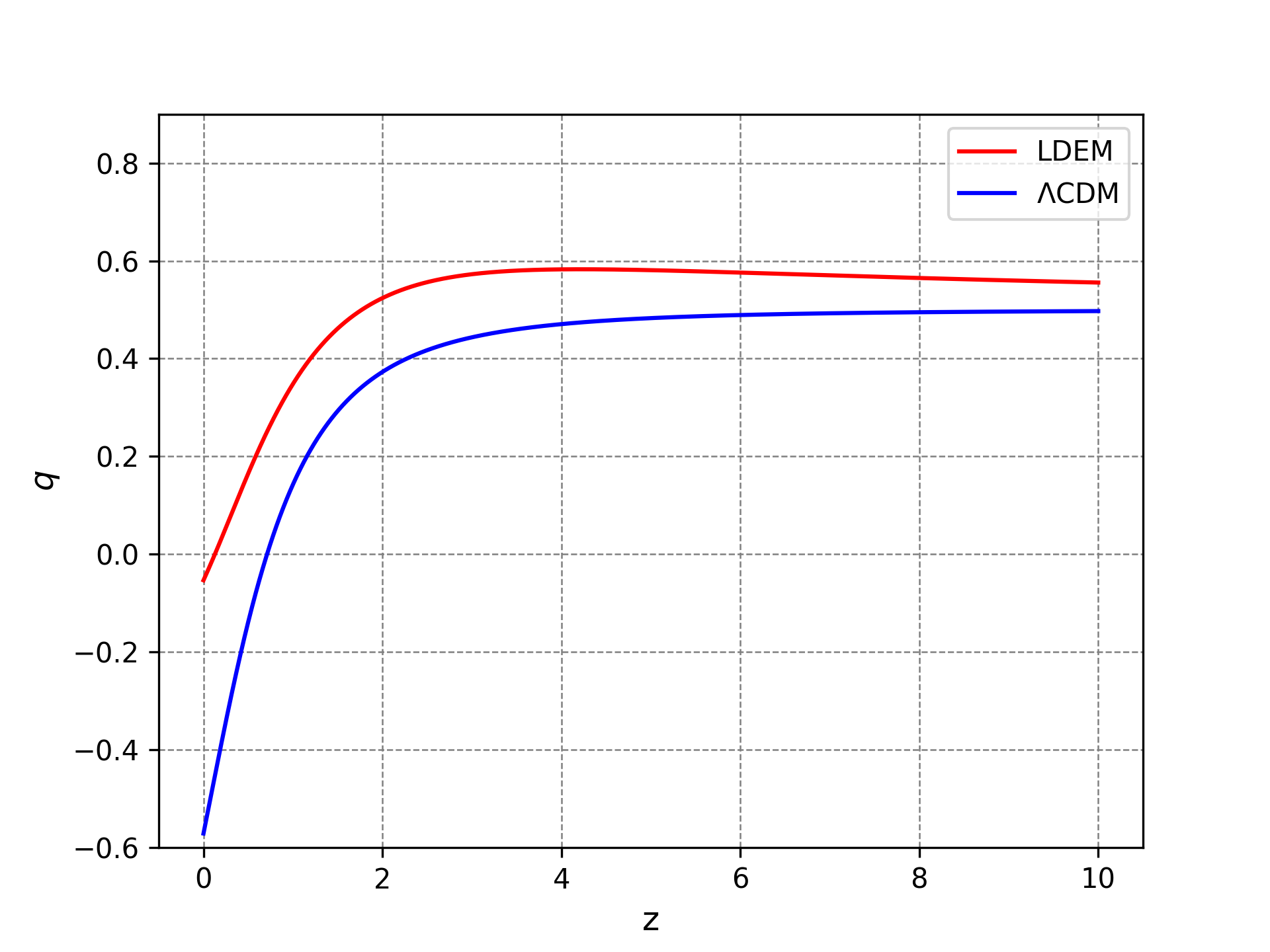}
	\caption{Evolution of the deceleration parameter ($q$) with redshift ($z$).}
	\label{fig:dec}
\end{figure}
On substituting the expression for Hubble parameter provided in Eq. (\ref{eq:H}), we obtain the evolution of deceleration parameter in LDEM model as,
\begin{align}
	\label{shift}
	q =  -1 + \frac{1}{2h^2}\left(2\Omega^{0}_{m}a^{-3}+\frac{\Omega^{0}_{m}(1+e)ae^{a}}{(1+e^a)^2}\right)
\end{align}
The evolution of the deceleration parameter against redshift is shown in Fig. (\ref{fig:dec}).
We found the present value of the deceleration parameter $(q_0)$ to be $-0.05$ for the LDEM, indicating that the present universe is in the accelerating phase. We also computed the transition redshift $(z_{T})$; indicating the redshift at which the universe switches from decelerating to an accelerated expansion to be $0.13$. The corresponding values of $q_0$ and $z_{T}$ for the $\Lambda$CDM model is estimated as $-0.55$ and $0.68$ respectively, showing a significant deviation from the LDEM model.
\section{Conclusion}
In this work, we introduced logistic dark energy model (LDEM) where the evolution of dark energy follows a logistic function with respect to the scale factor. This phenomenological model is inspired by previous research on quintessence-type scalar field models with Woods-Saxon-like potentials, which have successfully explained the late-phase acceleration of the universe\cite{radhakrishnan2024scalar}. The functional form for dark energy density that we considered frequently appears in physics as the Fermi-Dirac distribution and Woods-Saxon density distribution. Although there is no apparent connection between dark energy density and these distributions, their mathematical properties motivated us to explore this form for dark energy. Additionally, the simplicity of the logistic dark energy model, which possesses the same number of parameters as the $\Lambda$CDM model, makes it an attractive alternative. 

We tested the LDEM against observational data, including OHD, SNe Ia, BAO, and the CMB shift parameter. The results are compelling, with a $\Delta BIC$ of $-20.42$ for the OHD+SNe Ia combination and $-20.50$ for the OHD+SNe Ia+BAO combination, indicating strong evidence in favor of LDEM over the $\Lambda$CDM model. Interestingly, the Hubble constant obtained using any data combination aligns closely with the CMB prediction, effectively resolving the Hubble tension problem and suggesting a need to reconsider local measurements.

The cosmographic parameters obtained using the LDEM shows significant deviations from the standard model predictions. For instance, the LDEM model predicts a slightly higher age of the universe at 14.09 billion years, compared to the 13.85 billion years predicted by the $\Lambda$CDM model. Additionally, according to the LDEM scenario, the universe transitioned from deceleration to acceleration at a significantly lower redshift of 0.13, compared to 0.68 for the $\Lambda$CDM model.

The model is particularly intriguing because, despite its phenomenological nature, it shows strong evidence over the $\Lambda$CDM model. This indicates that the logistic function might capture essential aspects of the evolution of dark energy. Future research could further investigate the underlying physical mechanisms that could give rise to such a functional form for dark energy, potentially offering deeper insights into the nature of dark energy and the accelerating expansion of the universe. Additionally, further research at the perturbation level is needed to confirm these results. Exploring the effects of logistic dark energy in the early universe remains a promising area for future research.

\section*{Acknowledgment}
The authors of the manuscript are thankful to the Indian Institute of Technology Kanpur for providing the Param Sanganak high-performance computation facility for faster execution of the Python program. One of the authors, Sarath Nelleri, is thankful to the Indian Institute of Technology Kanpur for providing the Institute Postdoctoral Fellowship.
 

\end{document}